\documentclass[aps,superscriptaddress,twocolumn,preprintnumbers,amsmath,amssymb]{revtex4}
\usepackage{mathrsfs}
\usepackage{amsmath}
\usepackage{amssymb}
\usepackage{graphicx}
\usepackage{dcolumn}
\usepackage{bm}
\usepackage{subfigure}
\usepackage{color}
\usepackage{ifpdf}
\usepackage{epstopdf}
\usepackage{CJK}
\usepackage[
  pdfstartview=FitH,%
  bookmarks=true,%
  bookmarksopen=true,%
  breaklinks=true,%
  colorlinks=true,%
  linkcolor=blue,anchorcolor=blue,%
  citecolor=blue,filecolor=blue,%
  menucolor=blue,pagecolor=blue,%
  urlcolor=blue]{hyperref}
\usepackage{float}

\def\be{\begin{equation}}
\def\ee{\end{equation}}
\def\bea{\begin{eqnarray}}
\def\eea{\end{eqnarray}}
\def\nn{\nonumber}

\begin{document}
\begin{CJK*}{GBK}{song}

\title{Finite Temperature Phase Transition in a Cross-Dimensional Triangular Lattice}

\author{Shengjie Jin}
\affiliation{School of Electronics Engineering and Computer Science, Peking University, Beijing 100871, China}
\author{Xinxin Guo}
\affiliation{School of Electronics Engineering and Computer Science, Peking University, Beijing 100871, China}
\author{Peng Peng}
\affiliation{School of Electronics Engineering and Computer Science, Peking University, Beijing 100871, China}
\author{Xuzong Chen}
\affiliation{School of Electronics Engineering and Computer Science, Peking University, Beijing 100871, China}
\author{Xiaopeng Li}\email{xiaopeng\_li@fudan.edu.cn}
\affiliation{State Key Laboratory of Surface Physics, Institute of Nanoelectronics and Quantum Computing, and Department of Physics, Fudan University, Shanghai 200433, China}
\affiliation{Collaborative Innovation Center of Advanced Microstructures, Nanjing 210093, China}
\author{Xiaoji Zhou}\email{xjzhou@pku.edu.cn}
\affiliation{School of Electronics Engineering and Computer Science, Peking University, Beijing 100871, China}
\affiliation{Collaborative Innovation Center of Extreme Optics, Shanxi University, Taiyuan, Shanxi 030006, China}

\begin{abstract}
Atomic many-body phase transitions and quantum criticality have recently attracted much attention in non-standard optical lattices. Here we perform an experimental study of finite-temperature superfluid transition of bosonic atoms confined in a three dimensional triangular lattice, whose structure can be continuously deformed to dimensional crossover regions including  quasi-one and two dimensions. This non-standard lattice system provides a versatile platform to investigate many-body correlated phases.  For the three dimensional case, we find that the finite temperature superfluid transition agrees quantitatively with the  Gutzwiller mean field theory prediction, whereas tuning towards reduced dimensional cases, both quantum and thermal fluctuation effects are more dramatic, and the experimental measurement for the critical point becomes  strongly deviated from the mean field theory.  We characterize the fluctuation effects in the whole dimension crossover process. Our experimental results  imply  strong many-body correlations in the system beyond mean field description, paving a way to study quantum criticality near Mott-superfluid transition in finite temperature dimension-crossover lattices.
\end{abstract}


\maketitle
\end{CJK*}

\paragraph*{Introduction.---}
Phase transition, a ubiquitous concept in many-body physics, has long been a central subject in the study of condensed matter physics, describing a broad range of phenomena from superconductivity, magnetism, to Bose-Einstein condensation. With recent experimental developments, ultracold atoms in optical lattices have become a fascinating platform to explore quantum phase transitions with control and tuning capability unreachable in conventional systems~\cite{2008_Bloch_Dalibard_RMP,2015_Lewenstein_RPP}. Both fermionic and bosonic Hubbard models previously proposed as theoretical toy models to study strongly correlated physics in solid state systems, have now been precisely implemented by confining alkali atoms in optical lattices~\cite{Fisher,Jaksch,2002_Hofstetter_PRL,Greiner2002,2007_Lewenstein_AP,2008_Jordens_Nature,2008_Schneider_Science,2008_Bloch_Dalibard_RMP,2010_Esslinger_CMP,2015_Lewenstein_RPP,2016_Zohar_RPP,2016_Li_Liu_RPP,2017_Eckardt_RMP}. For the former, experimental efforts have been largely focused on the finite-temperature physics for the experimental challenge to reach the zero-temperature quantum ground states~\cite{2008_Jordens_Nature,2008_Schneider_Science,tarruell2012creating,2013_Esslinger_PRL,2015_Hulet_Nature,2016_Parsons_Science,2017_Greiner_Nature,2018_Bakr_NatPhys}. For the latter, the ground state superfluid phase and the Mott-superfluid transition, have been accomplished in optical lattices of different dimensionality and geometries~\cite{Greiner2002,2005_Esslinger_JLTP,sebby2006lattice,2011_Sengstock_Science,2011_Sengstock_NatPhys,wirth2011evidence,2012_Kurn_PRL,2012_Sengstock_NatPhys,2014_Sengstock_PRA}. It has been found that this phase transition is qualitatively captured by a Gutzwiller-type mean field theory~\cite{1991_Rokhsar_PRB}.

For a finite temperature system near a quantum phase transition point, it is well known that thermal fluctuations  play essential roles in characterizing the relevant physical properties, leading to a wide quantum critical region~\cite{2005_ColemanCriticality,sachdev2008quantum} potentially of deep connections to the understanding of high T$_c$ superconductivity~\cite{Bednorz1986} mechanism. Finite temperature effects near a Mott-superfluid transition have been explored in theory~\cite{Fisher,lu2006finite,kato2008sharp,zhou2010signature,Hazzard2011techniques,witczak2014dynamics}, but the experimental  studies are relatively scarce~\cite{2011_Chin_NJP,zhang2012observation,2017_Pan_PRL}. The temperature effects are particularly intricate  for optical lattices undergoing a dimension crossover, where thermal fluctuations intertwine with quantum kinematics.  For such systems, renormalization group analysis implies strong fluctuation effects and inapplicability of mean field theories even at a qualitative level~\cite{giamarchi2004quantum,zhao2008theory,lin2011u}. Characterizing quantum and thermal fluctuation effects beyond the mean field theory for the phase transition of an optical lattice in the dimension crossover region thus demands experimental studies.



\begin{figure*}[htp]
 \includegraphics[width=\textwidth]{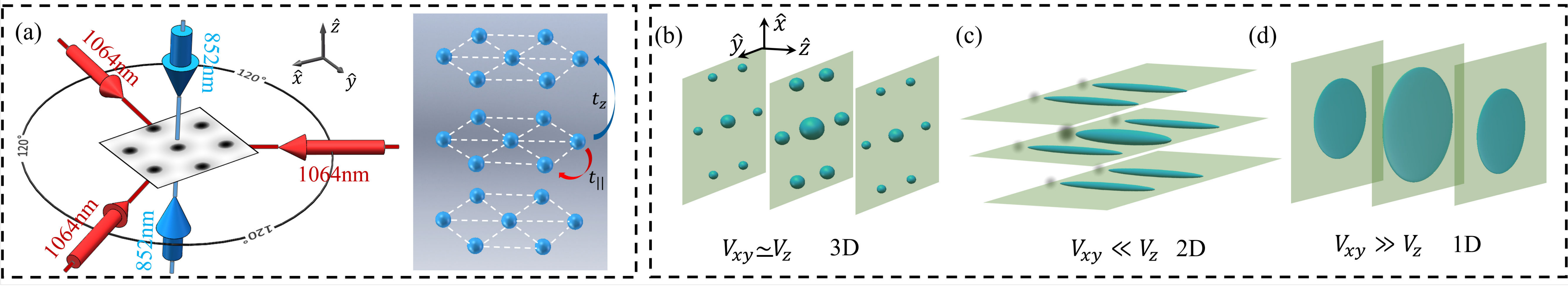}
  \caption{Pictorial illustration of the experimental lattice system. (a) shows the structure of the optical lattice and the spatial distribution of atoms. There are three laser beams in the $xy$-plane  with $\lambda_{\parallel}=1064$nm (red) forming a triangular pattern, and two laser beams in the $z$ direction with $\lambda_{z}=852$nm (blue) providing an additional one dimensional confinement. The blue and red arrows in the right panel of (a) represent the tunnelings $t_z$ and $t_{||}$, respectively. (b-d) Schematic diagram of momentum distribution. (b) corresponds to the three dimensional case with  $V_{xy}\simeq V_z$. (c) corresponds to the quasi-two dimensional case with  $V_{xy}\ll V_z$, where the momentum interference peaks are dispersed on $\hat{x}$-$\hat{y}$ plane. (d), quasi-one dimensional case with $V_{xy}\gg V_z$ where the interference are dispersed in $\hat{z}$ direction.
  }\label{fig:f1}
\end{figure*}

In this paper, we perform experimental studies of the finite-temperature superfluid phase transition in a triangular optical lattice whose dimensionality is continuously tuned from quasi-one dimension to two and three dimensions.
Atoms in lattices of different dimensionality can be clearly distinguished from the experimentally measured momentum distributions as probed in two orthogonal directions. As shown in Fig.~\ref{fig:f1}, the lattice contains a triangular lattice in the $xy$-plane and a one-dimensional lattice along $z$-axis.
 The lattice structure is determined by the two-dimensional triangular  lattice depth $V_{xy}$  and  the one dimensional lattice depth $V_z$. When $V_{xy}\simeq V_z$, the lattice is three dimensional, where we find phase transition properties agree with mean field theory predictions. The system becomes quasi-two and one-dimensional at $V_z\gg V_{xy}$, and $V_z\ll V_{xy}$, for both of which experimental measurements are strongly deviated from mean field theory predictions, yielding strong fluctuation effects in dimension crossover regions.
Our experiment paves a way to study novel many-body physics of dimension crossover lattices, where quantum and thermal fluctuations are both dominant.

\paragraph*{Experimental system and model description.---}

In our experiment, the triangular lattice is formed by three laser beams with a wavelength $\lambda_{\parallel} =1064$nm that intersect at the position of Bose-Einstein condensate (BEC) in $xy$-plane. For the confinement in the third direction, i.e., the $z$ axis, we add a vertical optical lattice formed by the interference of counter-propagating laser beams with a wavelength $\lambda_z = 852$nm.
The optical potential produced in this setup is then given by
\begin{eqnarray}
\label{Triangular Lattice}
&&V ({\bf x}) \\
&=& -V_{xy}\sum_{i=1}^{3}\cos(\frac{2\pi}{\lambda_{\parallel}} \sqrt{3}\vec{b}_i\cdot\vec{x}+\Delta \phi_i) +  V_z\cos^2(\frac{2\pi}{\lambda_z}  z), \nn
\end{eqnarray}
where
we have ${\bf  x} = (x,y,z)$, and $\vec{b}_1 = [1, 0, 0 ]$, $\vec{b}_2 = [-1/2, \sqrt{3}/2, 0]$, and $\vec{b}_3 = [-1/2, -\sqrt{3}/2, 0]$. Here the unit of potential $V_{xy}$ and $V_z$ is $E_R$, which equals to $\frac{\hbar k_{\parallel}^2}{2m}$ with $k_{\parallel}=\frac{2\pi}{\lambda_{\parallel}}$, respectively.

Before turning on the optical lattice, we prepare BECs of about $1.5\times10^5$ $^{87}$Rb atoms in a harmonic trap and the temperature is 50nK. The procedure details have been provided in our earlier works~\cite{NJP_zhou,PhysRevA.92.043614,PhysRevA.94.033624}. 
After the preparation of BECs, we adiabatically ramp on the triangular lattice within $80$ms.   Then the vertical lattice $V_z$ is adiabatically turned on within $20$ms.  Then we hold on the system for $10$ms. The average filling of the lattice is approximately six atoms per site. Finally, all the trap and lattices are turned off and an absorption image is obtained after time-of-flight. In the experiment, we can get the absorption image from  $z$-direction or $y$-direction, which is called Probe-Z and Probe-Y, respectively.

\begin{figure}
 \includegraphics[width=0.5\textwidth]{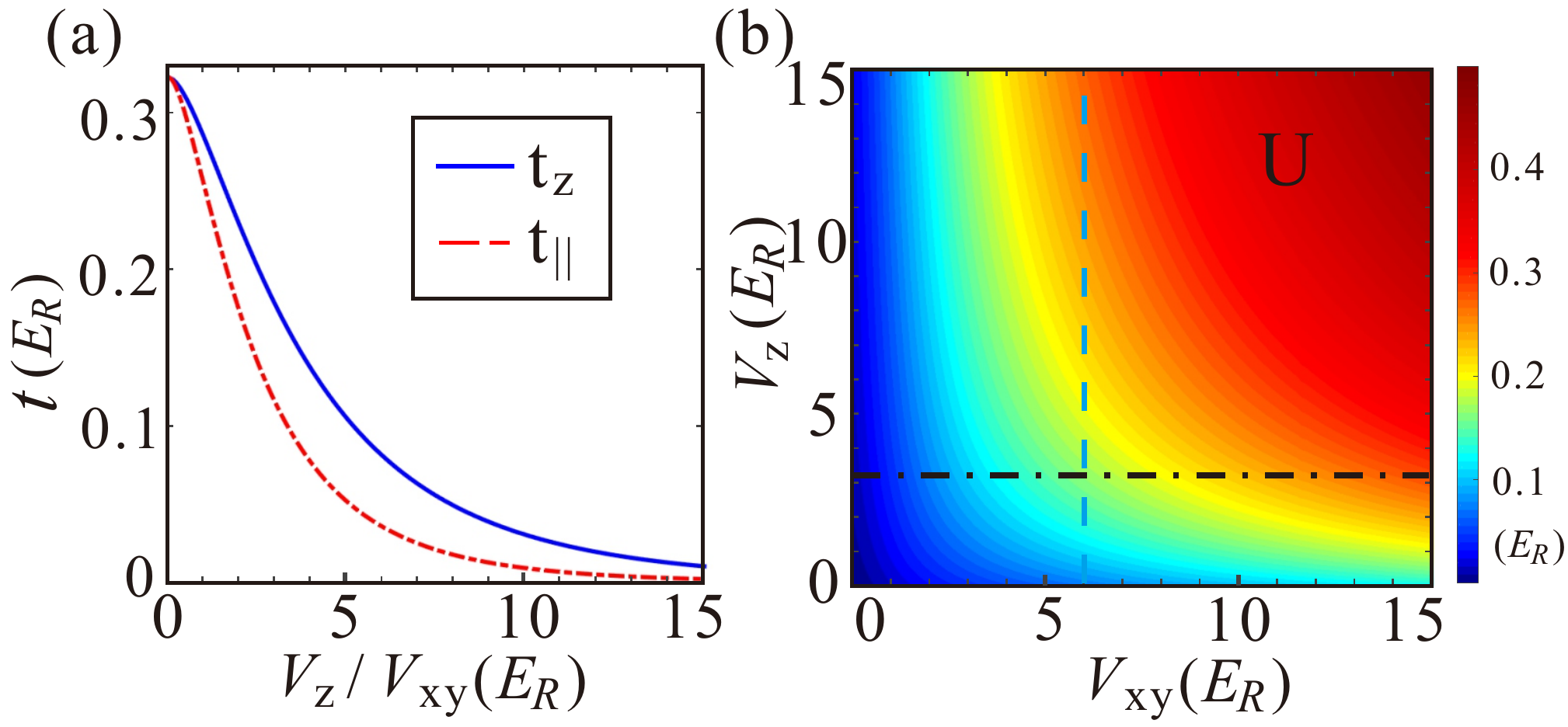}
  \caption{Parameters in the tight-binding model through exact band structure calculation. (a) shows the hopping amplitudes $t_{\parallel}$, and $t_z$ obtained by fitting the tight-binding energy dispersion to the band structure calculation. (b) The on-site interaction U obtained by a field theoretical tree level estimate (see Suppelementary Material). The ``dashed" and ``dash-dotted" lines correspond to $V_{xy} = 6E_R$ and $V_z = 3.2E_R$, for which the momentum distributions are shown in Fig.~\ref{fig:f2}.  
  }\label{fig:fig2}
\end{figure}

\begin{figure*}[htp]
 \includegraphics[width=\textwidth]{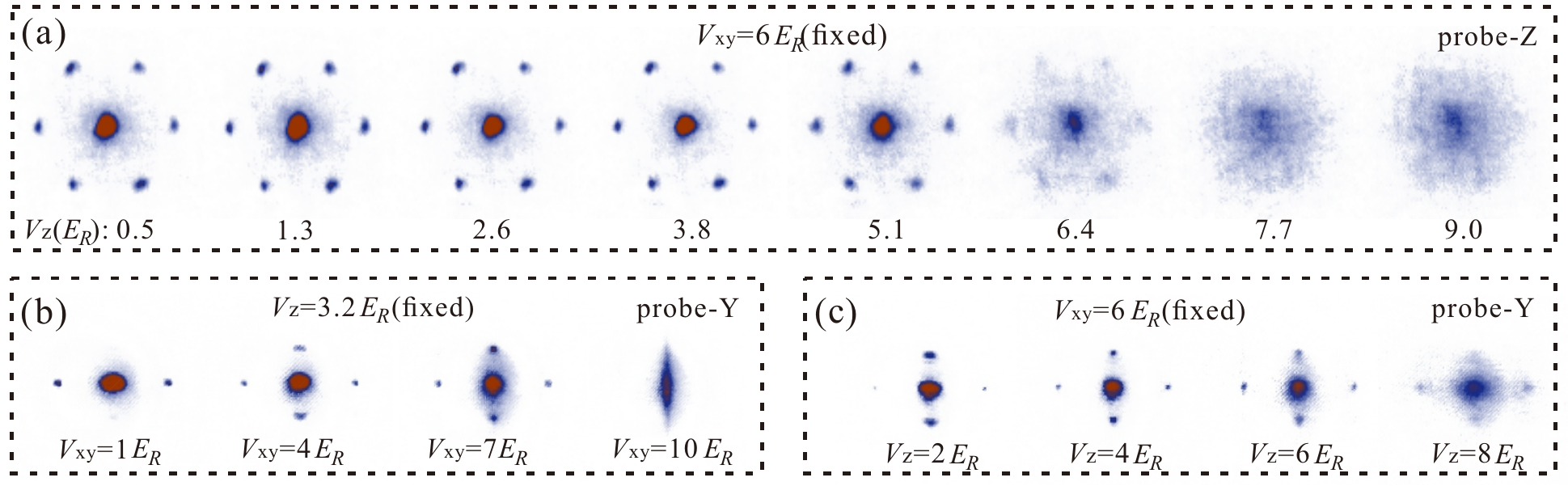}
  \caption{The superfluid to {normal bose liquid} transition observed in  different directions. (a) Observed in the $z$-direction (probe-Z), $V_{xy}$ is fixed at $6E_R$ and $V_z$ is from $0.5E_R$ to $9E_R$. (b) Observed in the $y$-direction (probe-Y), $V_z$ is fixed at $3.2E_R$ and $V_{xy}$ is from $1E_R$ to $10E_R$. (c) Observed in the $y$-direction (probe-Y), $V_{xy}$ is fixed at $6E_R$ and $V_z$ is from $2E_R$ to $8E_R$.
  }\label{fig:f2}
\end{figure*}

A single-band lattice Hamiltonian is reached under tight-banding approximation,
\begin{eqnarray} \label{Hamiltonian}
\begin{split}
H=&\sum_{<\textbf{r},\textbf{r}^{\prime}>}[-t_{||}\hat{b}_{\textbf{r}}^\dagger \hat{b}_{\textbf{r}^\prime}+H.c.]+\sum_{\textbf{r}}[-t_{z}\hat{b}_{\textbf{r}}^\dagger \hat{b}_{\textbf{r}+\textbf{e}_z}+H.c.]\\
 &+ \frac{U}{2}\sum_{\textbf{r}}\hat{b}_{\textbf{r}}^\dagger \hat{b}_{\textbf{r}}^\dagger \hat{b}_{\textbf{r}} \hat{b}_{\textbf{r}} -\mu\sum_\textbf{r} \hat{b}_{\textbf{r}}^\dagger \hat{b}_{\textbf{r}}.
 \end{split}
 \end{eqnarray}
Here $<{\bf r},{\bf r}^{\prime}>$ represents two neighboring sites in the $xy$-plane,  $\hat{b}$ and $\hat{b}^\dag$ the annihilation and creation operators, $U$ the interaction strength, $\mu$ the chemical potential. The tunnelings in $xy$-plane and in the $z$ direction are $t_{\parallel}$ and $t_z$, respectively (see Fig.~\ref{fig:f1}(a)). 
The tunnelings are determined by fitting the tight-binding energy dispersion to the band structure through exact calculations. The model parameters  as shown in Fig.~\ref{fig:fig2} are highly controllable by tuning  the lattice depths $V_{xy}$ and $V_z$ in our experimental setup.

\paragraph*{Dimensional crossover.---}
Since the lattice depths $V_{xy}$ and $V_z$ can be separately tuned in our experiment,  the dimensionality of the system is controllable. When $V_{xy}$ and $V_z$ are comparable, the system is a regular three dimensional lattice. In this region, mean field theory is expected to capture the essential physics, because  it is close to the upper critical dimension of the $U(1)$ phase transition. When $V_z$ is much weaker than $V_{xy}$, atoms are then less confined in the $z$ direction, and the system should be treated as weakly coupled one dimensional chains. The corresponding theoretical description is coupled Luttinger liquids, the transition temperature is determined by the inter-chain coupling~\cite{giamarchi2004quantum,zhao2008theory}. In the opposite limit, when $V_{xy}$ is much weaker than $V_z$, the system is formed of weakly coupled two dimensional layers, whose physical properties rely on the comparison between Kosterlitz Thouless transition temperature and inter-layer couplings~\cite{lin2011u}.
Then the superfluid transition in the dimension crossover regions should be taken as a finite-temperature phase transition rather than a zero temperature Mott-superfluid transition.

Fig.~\ref{fig:f2} shows time-of-flight measurements of the atomic system, which confirms our capability to control dimensionality from three- to dimension crossover regions. With the Probe-Z and -Y, we measured the momentum distribution in the $xy$- and $zx$- planes.  As shown in Fig.~\ref{fig:f2}(a,c), with decreasing $V_z$ at a fixed $V_{xy}$, $t_z$ becomes larger, which drives a phase transition from normal to a superfluid. In the small $V_z$ limit, the system behaves as a coupled array of Luttinger liquids.
Fig.~\ref{fig:f2} (b) corresponds to varying $V_{xy}$ with a fixed $V_z$. In this case, the system is formed of weakly coupled two dimensional systems at small $V_{xy}$. The weakening of phase coherence with larger $V_{xy}$ is attributed to the decrease in the tunneling $t_\parallel$.

\paragraph*{Finite temperature mean field theory.---}
To characterize fluctuation and many-body correlation effects beyond mean field theory in the dimension crossover regions,
we provide a finite temperature mean field theory to compare with experimental results.
Under the mean field approximation, the density matrix of the system is given by
$\rho \propto \prod_{\bf r} e^{-\beta H_M ({\bf r})  } $, with 
$\beta$ the inverse temperature,  $H_M  ({\bf r})  =-t_{\rm eff} \left( \varphi \hat{b}_{\bf r} ^\dag+\varphi^* \hat{b}_{\bf r}   \right)/2   + U\hat{n}_{\bf r}(\hat{n}_{\bf r}-1)/2 - \mu \hat{n}_{\bf r} $,  the tunneling parameter
$t_{\rm eff} = 6 t_\parallel + 2 t_z$, and  the superfluid order parameter  $\varphi = \left[ {\rm Tr} e^{-H_M ({\bf r} )}  \hat{b}_{\bf r}  \right]/ \left[ {\rm Tr} e^{-H_M ({\bf r})} \right] $, which is self-consistently determined. The mean field phase diagram is solely dependent on  two dimensionless parameters, $ k_B T/t_{\rm eff}$ and $U/t_{\rm eff}$, which characterize the strengths of thermal  and quantum fluctuations.
At the zero temperature limit, an instability analysis shows Mott-superfluid phase boundary is given by
$
t_{\rm eff} ^c = -\left[ \mu + (1- n)U\right] \left[\mu - nU\right]/\left[\mu +U\right],
$
with $n$ the filling of the Mott state. Considering a transition with a fixed particle number, the phase boundary is further reduced to
$t_{\rm eff} ^c/U = 2n+1 - \sqrt{(2n+1)^2 -1}$, reproducing the previous ground state analysis for symmetric lattices~\cite{Fisher,Jaksch}.  At finite temperature, we rely on numerical self-consistent calculations to compare with experimental results.

\begin{figure}
 \includegraphics[width=0.5\textwidth]{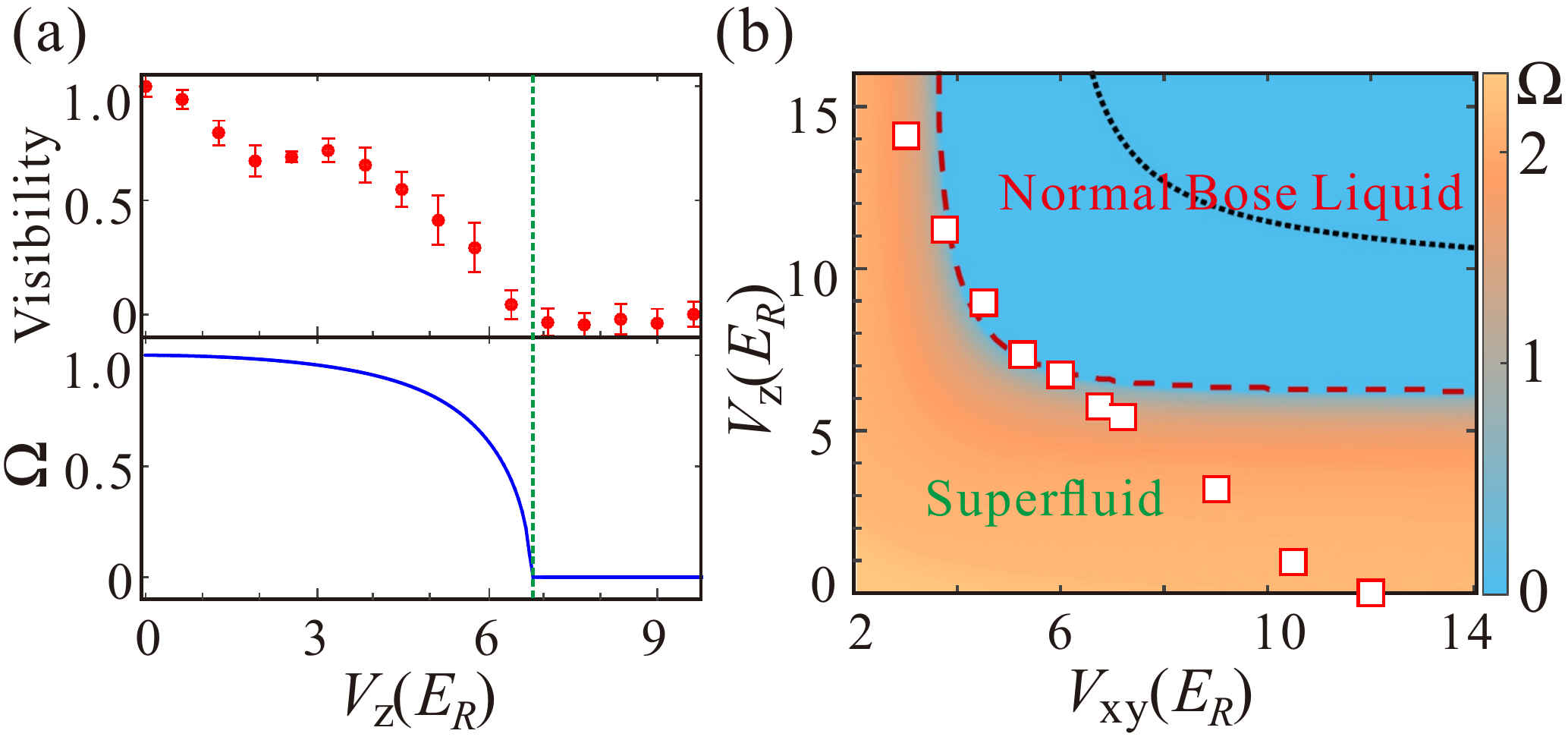}
  \caption{The superfluid transition.
 (a), signature of the superfluid transition from the visibility. The blue lines are the superfluid order parameters calculated by mean field theory. The red points are the visibility measured in experiment. The in-plane lattice potential is fixed at   $V_{xy}=6E_R$. Both visibility and the theoretical superfluid order parameter are normalized with respect to the $V_z \to 0$ limit. (b) The superfluid phase diagram. The blue and brown areas represent the high temperature normal bose liquid and the superfluid phases, respectively, obtained from mean field theory.  {The dashed line represents the phase boundary obtained from finite temperature Gutzwiller mean field theory. The dotted line representing the zero temperature mean-field Mott-superfluid phase boundary as a comparison.} The red squares represent the transition points measured in experiments.
  }\label{fig:f3b}
\end{figure}

\paragraph*{Experimental determination of finite temperature phase diagram.---}
We perform experimental measurements of the superfluid transition in different parameter regions  of the system corresponding to three-, quasi-one and two dimensions.  Firstly, we increase $V_z$ with a fixed $V_{xy}$. As shown in Fig.~\ref{fig:f2}(a), we change $V_z$ from zero to 9$E_R$ with $V_{xy}$ fixed at $6E_R$. The superfluid phase transition is revealed with Probe-Z. There are several methods to get the transition point from the absorption images~\cite{Greiner2002,Gerbier,Yi,Spielman,Becker}. Here we choose the visibility to determine the transition point (see Supplementary Material)~\cite{Gerbier,Becker}.
The visibility of the time-of-flight images across the transition is shown in Fig.~\ref{fig:f3b}(a), according to which the transition point is determined in the experiment. In Fig.~\ref{fig:f3b}(a),
{the transition point is  $V_z=6.7E_R$}. According to the finite temperature mean field theory and the parameters in Fig.~\ref{fig:fig2}, we can get the superfluid order parameters in theory, which is shown in Fig.~\ref{fig:f3b}(a) by the blue line.
{The theoretical superfluid order parameters reduce to zero at $V_z=6.8E_R$}.
The experimental measurement thus agrees quantitatively with mean field theory prediction.

\begin{figure}
 \includegraphics[width=0.5\textwidth]{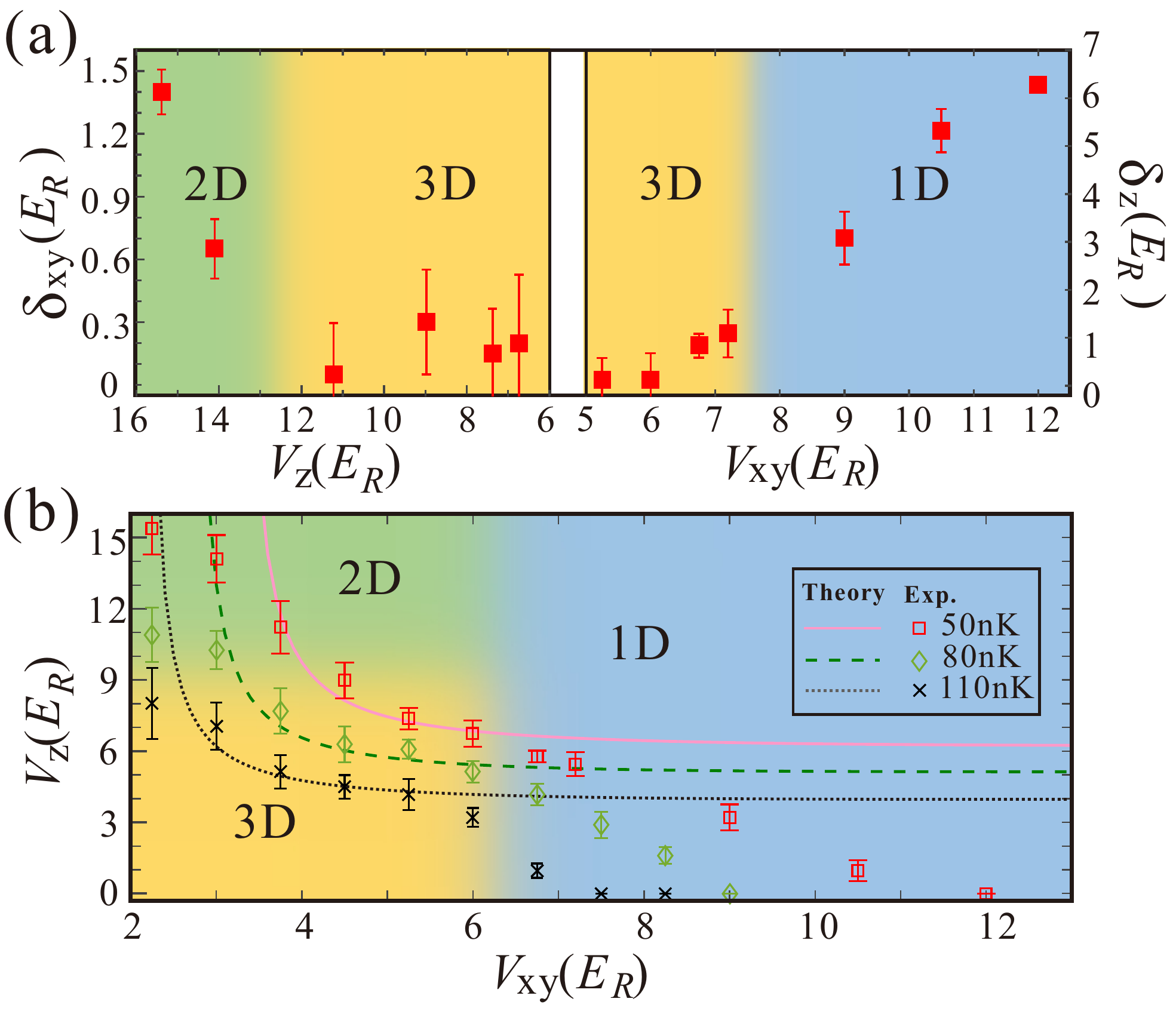}
  \caption{ (a) The discrepancy in  the transition point by comparison between experimental measurements and mean field theory prediction. In the left panel, $\delta_{xy}$ is the difference in the critical value of $V_{xy}$ between experimental measurement and mean field theory, when $V_z$  is fixed, which systematically increases as we approach the two dimensional limit from the three dimensional case. In the right panel, $\delta_{z}$ is the difference in the critical value of $V_z$ when $V_{xy}$ is fixed, which also systematically increases as we approach the one dimensional limit.
  (b) The phase diagram obtained from theory calculation and experiment. The red, green and black represent the results of 50nK, 80nK, and 110nK, respectively. The lines are from theory calculations and the points are from experiment.
  }\label{fig:f4}
\end{figure}

Choosing different $V_{xy}$, we can get critical strengths of $V_z$ for the superfluid transition as a function of $V_{xy}$.
Then we get  a finite temperature phase diagram as shown in Fig.~\ref{fig:f3b}(b), where  the  measured transition points are represented by the red squares. According to the finite temperature mean field theory and the parameters obtained in Fig.~\ref{fig:fig2}, we can get the phase diagram in theory. As shown in Fig.~\ref{fig:f3b}(b), the blue area represents the bosonic normal fluid phase  in theory and the brown area represents the superfluid phase. The transition boundary at the zero temperature limit is also shown by a black dotted line in Fig.~\ref{fig:f3b}(b).  For the three dimensional case, with $V_{xy}$ comparable with $V_z$,  the measured phase transition point  agrees with the theory, whereas for the quasi-two and one dimensional cases corresponding to Fig.~\ref{fig:f1}(d) with $V_{xy}\ll V_z$ and  Fig.~\ref{fig:f1}(c) with $V_{xy}\gg V_z$, we find significant deviation of the experimental measurement from mean field theory prediction.  The breakdown of mean field theory in dimension crossover regions is due to strong fluctuation or  many-body correlation effects neglected in the mean field theory. We characterize the fluctuation effects by the difference of measured superfluid transition point from the theory prediction, in $V_{xy}$ with fixed $V_z$ and in $V_z$ with fixed $V_{xy}$ for the quasi-two and quasi-one dimensional cases, respectively.
The results are shown in  Fig.~\ref{fig:f4}(a). The systymatic increase of fluctuation effects as we go from the three to one or two dimensions are clearly revealed.

In the above discussion, the temperature of the quantum gas is 50nK. We also  systematically study the temperature effect on the superfluid transition phase boundary.  The lines in Fig.~\ref{fig:f4} show the transition points for different temperatures, 50nK, 80nK, and 110nK. For the three temperatures, the number density  of the atomic gas remains unchanged. For these different temperatures, we  use the same method as in Fig.~\ref{fig:f3b} to find the transition points. For instance, if we fix $V_{xy}=6E_R$, the transition points are $V_z=6.7E_R$, $5.1E_R$, and $3.2E_R$ corresponding to the temperature 50nK, 80nK and 110nK, respectively. The theoretical results are shown in Fig.~\ref{fig:f4}(b). At these different temperatures, we still see the experimental agreement (disagreement) in the three dimensional (quasi-two and one dimensional) case. The major observed effect of increasing temperature is the decrease in the critical lattice potential. The significance of strong quantum and thermal fluctuations are revealed in the experiment, which implies the dimension crossover lattices provide a natural platform to study quantum critical behaviors, of fundamental interest to the understanding of high Tc superconductivity in cuperates.

\paragraph*{Conclusion.---}
To conclude, we studied the finite-temperature superfluid transition in a three dimensional triangular lattice, continuously tuned from three to quasi-one and two dimensions.  For the three dimensional case, the experimentally measured superfluid transition point is found to agree with the  Gutzwiller mean field theory prediction, whereas it strongly deviates from the mean field theory in the reduced dimensional cases, revealing strong many-body correlation effects in this optical lattice system. The strong quantum and thermal fluctuation effects established in the dimension crossover regions of our triangular optical lattice, suggest rich quantum critical behavior worth further theoretical and experimental exploration.

\paragraph*{Acknowledgement.}

This work is supported by National Program on Key Basic Research Project of China (Grant No. 2016YFA0301501, Grant No. 2017YFA0304204),
and National Natural Science Foundation of China (Grants No.11334001, and No. 61727819, and No. 117740067).
XL also acknowledges support by the Thousand-Youth-Talent Program of China.

\bibliographystyle{apsrev4-1}
\bibliography{my}

\begin{widetext}
\newpage 

\begin{center} 
{\bf \Huge Supplementary Material}
\end{center} 
\renewcommand{\theequation}{S\arabic{equation}}
\renewcommand{\thesection}{\large S-\arabic{section}}
\renewcommand{\thefigure}{S\arabic{figure}}
\renewcommand{\bibnumfmt}[1]{[S#1]}
\renewcommand{\citenumfont}[1]{S#1}
\setcounter{equation}{0}
\setcounter{figure}{0}


\begin{figure}[htp]
 \includegraphics[width=0.4\textwidth]{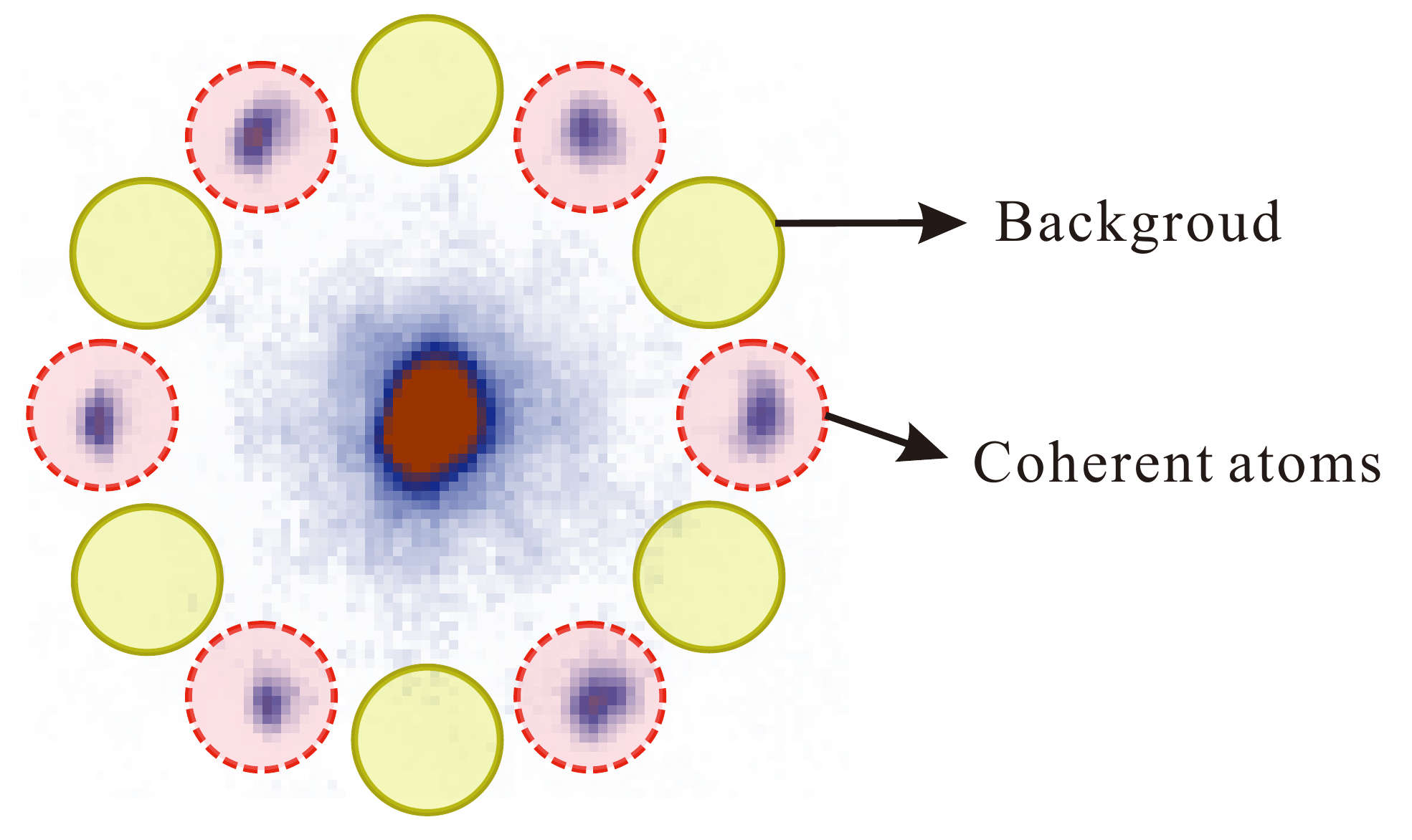}
  \caption{ The extraction of visibility. In this absorption image, the six red areas represent the positions of the first order momentum states. The yellow areas represent the background position at the edge of the first Brillouin zone.
  }\label{fig:visibility}
\end{figure}

\section{Visibility}

The visibility can be extracted from the absorption images as shown in Fig.~\ref{fig:visibility}. We count the number of atoms in the red areas, which indicates the atoms in first order momentum states, and the total number of atoms in these six red areas is expressed as $N_{atoms}$. In the contract, we also count the atoms in the yellow areas, which indicates the incoherent atoms at the edge of the first Brillouin zone and expressed as $N_{background}$. Then, the visibility is given as:
\begin{eqnarray}
Visibility=\frac{N_{atoms}-N_{background}}{N_{atoms}+N_{background}}.
\end{eqnarray}
It should be pointed out that the visibility in the main text is normalized. And the absolute value of visibility at $V_{z}=0$ is about 0.45.

\section{Hamiltionian}
The corresponding model description for our system can be derived from the theory in continuum, with a Hamiltonian
\be
H =\textstyle \int d^3  {\bf x} \left\{ \phi^\dag ({\bf x}) \left [ -\frac{\hbar^2 \vec{\nabla}^2} {2M} + V({\bf  x})  -\mu\right] \phi ({\bf  x}) + g \phi^\dag \phi^\dag \phi \phi \right\},   \nn
\ee
with $\phi({\bf  x})$ a bosonic field operator, $M$ the atomic mass,  $\mu$ the chemical potential, and $g = 2\pi\hbar^2 a_s /M$ ($a_s$ is the s-wave scattering length) the interaction strength. We expand the field in terms of Wannier basis as $\phi({\vec x}) = \sum_{\bf r} w ({\bf  x} - {\bf r}) b_{\bf r} $,  with $w({\bf x} - {\bf r})$ the localized Wannier function, and $b_{\bf r}$ the associated lattice annihilation operator.  A single-band lattice Hamiltonian is then reached under tight-banding approximation,
\begin{eqnarray} \label{Hamiltonian}
\begin{split}
H=&\sum_{<\textbf{r},\textbf{r}^{\prime}>}[-t_{||}b_{\textbf{r}}^\dagger b_{\textbf{r}^\prime}+H.c.]+\sum_{\textbf{r}}[-t_{z}b_{\textbf{r}}^\dagger b_{\textbf{r}+\textbf{e}_z}+H.c.]\\
 &+ \frac{U}{2}\sum_{\textbf{r}}b_{\textbf{r}}^\dagger b_{\textbf{r}}^\dagger b_{\textbf{r}} b_{\textbf{r}} -\mu\sum_\textbf{r}b_{\textbf{r}}^\dagger b_{\textbf{r}}.
 \end{split}
 \end{eqnarray}
The tunneling parameters are then calculated by fitting to the exact band structure, and the interaction $U$ is estimated under field theoretical tree level approximation as 
\be 
U = g \int d^3 {\bf x} |w({\bf x})|^4 .
\ee

\section{Mean field order parameter} 

In the main text, we can see that when $V_z$ is very small, the transition points are no longer reduced as the depth $V_{xy}$ becomes larger. The reason of this phenomenon is because the $t_{\rm eff}=t_{\parallel}+t_z$ is determined by $t_z$ and remains the same as $V_{xy}$ increasing when $V_z$ is small(see Fig.~\ref{fig:all}(a)). At the same time, as $V_{xy}$ increasing, the chemical potential $\mu$ and interaction $U$ are increasing together, with their ratio roughly unaffected.  As a consequence, for small $V_z$, the order parameters will not reduce to zero as $V_{xy}$ increasing.

\begin{figure}[htp]
 \includegraphics[width=0.7\textwidth]{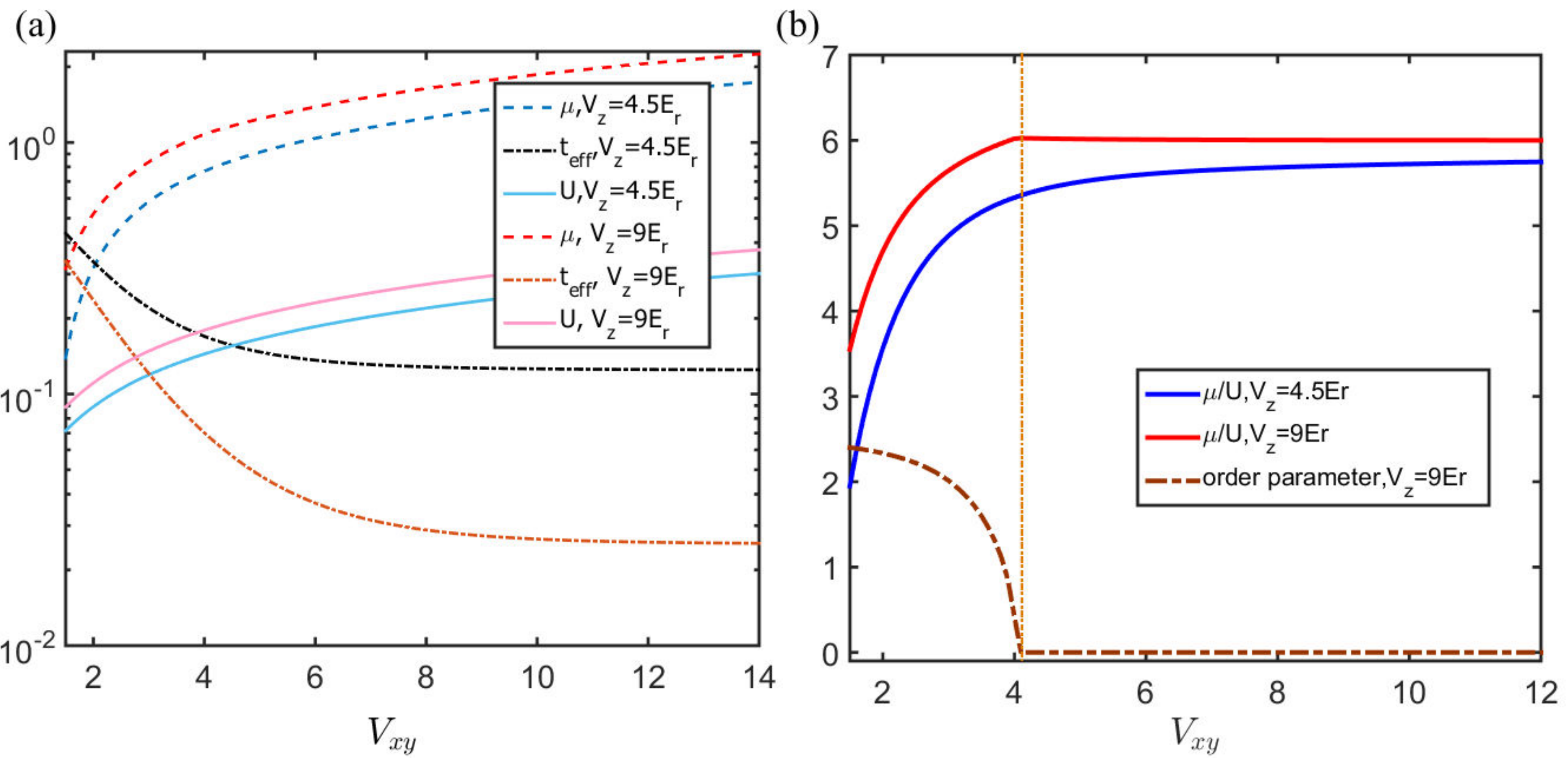}
  \caption{ The parameters in Bose-Hubbard model for $V_z=4.5E_R$ and $9E_R$. (a) The dashed lines, dash-dot lines and solid lines represent the chemical potential $\mu$, hopping amplitude $t_{\rm eff}$ and on-site interaction $U$, respectively. (b) The ratio of chemical potential $\mu$ to interaction $U$ and the order parameter for $V_z=9E_R$.
  }\label{fig:all}
\end{figure}

\end{widetext}

\end{document}